\documentclass[12pt]{article}
\usepackage{amsmath}
\usepackage{amssymb}
\usepackage{amsthm}
\usepackage{fullpage}
\usepackage[utf8]{inputenc}
\usepackage{mathtools}
\usepackage{url}
\usepackage[usenames,svgnames]{xcolor}

\usepackage[pagebackref]{hyperref} %hyperref wants to be loaded last
\usepackage[all]{hypcap} %fix hyperref links to tables/figures/etc.
\theoremstyle{plain}

\theoremstyle{definition}

\numberwithin{equation}{section} % requires package amsthm

\hypersetup{
  breaklinks,colorlinks,
  citecolor=Green,
  linkcolor=BlueViolet,
  urlcolor=DarkBlue,
  pdftitle={On  some class of partial difference equations admitting a zero-curvature representation},
  pdfauthor={Svinin Andrei K.},
}

\begin{document}
\baselineskip 16pt

\medskip
\begin{center}
\begin{Large}\fontfamily{cmss}
\fontsize{17pt}{27pt}
\selectfont
\textrm{On  some class of partial difference equations admitting a zero-curvature representation}
\end{Large}\\
\bigskip
\begin{large} {Svinin Andrei K.}
 \end{large}
\\
\bigskip
\begin{small}
{\em Institute for System Dynamics and Control Theory\\
Siberian Branch of Russian Academy of Sciences \\}
svinin@icc.ru \\
\end{small}
\end{center}
\bigskip

\begin{abstract}
We show some classes of higher order partial difference  equations admitting a zero-curvature representation and generalizing lattice potential KdV equation. We construct integrable hierarchies which, as we suppose, yield generalized symmetries for obtained class of partial difference  equations. As a byproduct we also derive non-evolutionary differential-difference equations with their Lax pair representation  which may be of potential interest.
\end{abstract}

%\keywords{integrable lattices; integrable hierarchy; zero-curvature representation}
%\ccode{2000 Mathematics Subject Classification: 37K10 }

%%%%%%%%%%%%%%%%%%%%%%%%%%%%%%%%%%%%%%%%%%%%%%%%%%%%%%%%%%%%%%%%%%%%%%%%%%%%%%%%%%%%%%%%%%%%%%%%%%%%%%%%%%%%%%%%%%%%%%%%%%%%%%%%%%%%%%%%%%%%%%%%%%%%%%%%%%%%%%%%%%%%%%%%%%%%%%%%%%%%%%%
\section{Introduction}
%%%%%%%%%%%%%%%%%%%%%%%%%%%%%%%%%%%%%%%%%%%%%%%%%%%%%%%%%%%%%%%%%%%%%%%%%%%%%%%%%%%%%%%%%%%%%%%%%%%%%%%%%%%%%%%%%%%%%%%%%%%%%%%%%%%%%%%%%%%%%%%%%%%%%%%%%%%%%%%%%%%%%%%%%%%%%%%%%%%%%%%%

The simplest case of an evolutionary differential-difference equation sharing the property of having Lax pair representation  is given by the Volterra lattice \cite{Volterra}
\begin{equation}
T_i^{\prime}=T_i(T_{i-1}-T_{i+1}).
\label{Vl}
\end{equation}
Integrability properties of this equation were first investigated in \cite{Manakov}, \cite{Hirota}. One knows that equation (\ref{Vl}) is related via the substitution $T_i=u_iu_{i+1}$ to its modified version
\begin{equation}
u_i^{\prime}=u_i^2(u_{i-1}-u_{i+1}).
\label{mVl}
\end{equation}
In turn, the substitution $T_i=1/(v_{i}-v_{i-2})(v_{i+1}-v_{i-1})$ gives the relationship of the Volterra lattice (\ref{Vl}) with one of an equation of Volterra type \cite{Yamilov}
\begin{equation}
v_i^{\prime}=\frac{1}{v_{i+1}-v_{i-1}}.
\label{yVl}
\end{equation}
As is known,  this equation plays the role of generalized symmetry for lattice potential KdV (lpKdV) equation\footnote{Note that this equation in fact are equivalent to partial difference one if one introduces additional variable $j$ and further identifies $v_i=v_{i, j}$ and $\bar{v}_i=v_{i, j+1}$.} 
\begin{equation}
\left(\bar{v}_i-v_{i+1}\right)(\bar{v}_{i+1}-v_i)=c.
\label{lpkdV}
\end{equation}
Another integrability property of this equation is its zero-curvature representation \cite{Adler}. In the paper we present a class of higher order equations of the form 
\begin{equation}
Q(v_i,\ldots, v_{i+p-1}, \bar{v}_i,\ldots, \bar{v}_{i+p-1})=0
\label{form}
\end{equation}
parametrized by pairs of co-prime numbers $(n, h)$. These equations in a sense generalize (\ref{lpkdV}). It should be noted that, recently, equations of the form (\ref{form}) were considered in \cite{Adler1} from the point of view of three-dimensional consistency and some examples were given there.  Partial difference equations presented below share a property of having a zero-curvature representation which constructed as Darboux transformation for some linear problem. Also we construct integrable hierarchies of evolution equations on the field $v_i$ which appear as compatibility conditions of the linear problem with corresponding evolution linear equations. It gives us hope that these evolution equations are generalized symmetries for corresponding partial difference ones.

%All the equations (\ref{Vl}), (\ref{mVl}) and (\ref{yVl}) are evolutionary ones. Moreover these equations admit corresponding hierarchies of generalized symmetries \cite{Nagai} which can be presented in explicit form via some discrete polynomials \cite{Svinin1}, \cite{Svinin2}. It is common of knowledge that  equations (\ref{Vl}), (\ref{mVl}) and (\ref{yVl}) have corresponding integrable generalizations. For example, the equation
%\[
%T_i^{\prime}=T_i\left(\sum_{j=1}^nT_{i-j}-\sum_{j=1}^nT_{i+j}\right),
%\]
%known as Itoh-Narita-Bogoyavlenskii lattice \cite{Itoh}, \cite{Narita}, \cite{Bogoyavlenskii}, naturally generalizes the Volterra lattice (\ref{Vl}), while its modified version looks as \cite{Bogoyavlenskii}
%\[
%u_i^{\prime}
%=u_i^2\left(\prod_{j=1}^nu_{i-j}-\prod_{j=1}^nu_{i+j}\right).
%\]
%Corresponding integrable generalization of equation (\ref{yVl}), namely,
%\begin{equation}
%\displaystyle
%v_{i}^{\prime}=\frac{1}{\prod_{j=1}^n\left(v_{i-j+n+1}-v_{i-j}\right)}
%\label{gen}
%\end{equation}
%and explicit form of its hierarchy was given in \cite{Svinin2}. In the paper we give further generalization of these equations.

A rest of the paper is organized as follows. In section \ref{sec:2}, we consider some class of auxiliary linear equations and look for compatibility conditions for these ones. In a result, we get differential-difference equations in a sense admitting Lax pair representation. It is worth remarking that resulting equations, generally speaking, are not of evolutionary type. In section \ref{sec:3}, we consider Darboux transformation of auxiliary linear equations and derive some class of quadratic partial difference equations as a condition of compatibility of corresponding linear discrete equations.  In section \ref{sec:4} we derive some integrable hierarchies which, as we suppose, are generalized symmetetries for corresponding partial difference equation of lpKdV type.  

%%%%%%%%%%%%%%%%%%%%%%%%%%%%%%%%%%%%%%%%%%%%%%%%%%%%%%%%%%%%%%%%%%%%%%%%%%%%%%%%%%%%%%%%%%%%%%%%%%%%%%%%%%%%%%%%%%%%%%%%%%%%%%%%%%%%%%%%%%%%%%%%%%%%%%%%%%%%%%%%%%%%%%%%%%%%%%%%%%%%%%%%%%%%%%%%%%
\section{Linear equations and its consistency conditions}
%%%%%%%%%%%%%%%%%%%%%%%%%%%%%%%%%%%%%%%%%%%%%%%%%%%%%%%%%%%%%%%%%%%%%%%%%%%%%%%%%%%%%%%%%%%%%%%%%%%%%%%%%%%%%%%%%%%%%%%%%%%%%%%%%%%%%%%%%%%%%%%%%%%%%%%%%%%%%%%%%%%%%%%%%%%%%%%%%%%%%%%%%%%%%%%%%%%

\label{sec:2}

%%%%%%%%%%%%%%%%%%%%%%%%%%%%%%%%%%%%%%%%%%%%%%%%%%%%%%%%%%%%%%%%%%%%%%%%%%%%%%%%%%%%%%%%%%%%%%%%%%%%%%%%%%%%%%%%%%%%%%%%%%%%%%%%%%%%%%%%%%%%%%%%%%%%%%%%%%%%%%%%%%%%%%%%%%%%%%%%%%%%%%%%%%%%%%%%%%%
\subsection{The first class of differential-difference equations}
%%%%%%%%%%%%%%%%%%%%%%%%%%%%%%%%%%%%%%%%%%%%%%%%%%%%%%%%%%%%%%%%%%%%%%%%%%%%%%%%%%%%%%%%%%%%%%%%%%%%%%%%%%%%%%%%%%%%%%%%%%%%%%%%%%%%%%%%%%%%%%%%%%%%%%%%%%%%%%%%%%%%%%%%%%%%%%%%%%%%%%%%%%%%%%%%%%%

Let us consider the following pair of linear equations:
\begin{equation}
zs_{i+n}\phi_{i+n}+\phi_i=z\phi_{i+n+h},\;\;\;
\phi_i^{\prime}=z\xi_{i}\phi_{i+n},
\label{linear1}
\end{equation}
on some wave function $\phi=\phi_i=\phi_i(x, z)$. By assumption, they are parameterized by some positive integers $n$ and $h$. As is seen, these equations constitute compatible pair provided that two relations, namely,
\begin{equation}
s_i\xi_{i}=s_{i+n}\xi_{i+h},\;\;\;
s_i^{\prime}=\xi_{i+h}-\xi_{i-n}
\label{pair}
\end{equation}
are valid. Remark that the first relation in (\ref{pair}) can be equivalently rewritten as
\begin{equation}
\prod_{j=1}^{h}\xi_{i+j-1}\prod_{j=1}^ns_{i+j-1}=\delta
\label{1}
\end{equation}
with some arbitrary constant $\delta$. By suitable reparametrization, we can make $\delta =1$. It is a simple observation that putting
\begin{equation}
\xi_i=\prod_{j=1}^nu_{i+j-1},\;\;\;
s_i=\prod_{j=1}^{h}\frac{1}{u_{i+j-1}},
\label{xiu}
\end{equation}
with some field $u=u_i$, we solve (\ref{1}) with $\delta=1$.  This ansatz, after substituting it in the second relation in (\ref{pair}), gives differential-difference equation
\begin{equation}
\left(\prod_{j=1}^{h}\frac{1}{u_{i+j-1}}\right)^{\prime}=\prod_{j=1}^nu_{i+j+h-1}-\prod_{j=1}^nu_{i+j-n-1}
\label{u1}
\end{equation}
which we can rewrite as
\begin{equation}
\sum_{j=1}^{h}\frac{u_{i+j-1}^{\prime}}{u_{i+j-1}}=\prod_{j=1}^{n+h}u_{i+j-n-1}-\prod_{j=1}^{n+h}u_{i+j-1}
\label{u2}
\end{equation}
or in the form
\begin{eqnarray}
\left(\prod_{j=1}^{h}u_{i+j-1}\right)^{\prime}&=&\prod_{j=1}^{h}u_{i+j-1}\left(\prod_{j=1}^{n+h}u_{i+j-n-1}-\prod_{j=1}^{n+h}u_{i+j-1}\right) \nonumber \\
&=&\prod_{j=1}^{h}u_{i+j-1}^2\left(\prod_{j=1}^nu_{i+j-n-1}-\prod_{j=1}^nu_{i+j+h-1}\right).
\label{u}
\end{eqnarray}
Remark that relations (\ref{u1}) and (\ref{u2}) can be considered as differential-difference conservation laws for equation (\ref{u}).

%%%%%%%%%%%%%%%%%%%%%%%%%%%%%%%%%%%%%%%%%%%%%%%%%%%%%%%%%%%%%%%%%%%%%%%%%%%%%%%%%%%%%%%%%%%%%%%%%%%%%%%%%%%%%%%%%%%%%%%%%%%%%%%%%%%%%%%%%%%%%%%%%%%%%%%%%%%%%%%%%%%%%%%%%%%%%%%%%%%%%%%%%%%%%%%%%
\subsection{The second class of differential-difference equations}
%%%%%%%%%%%%%%%%%%%%%%%%%%%%%%%%%%%%%%%%%%%%%%%%%%%%%%%%%%%%%%%%%%%%%%%%%%%%%%%%%%%%%%%%%%%%%%%%%%%%%%%%%%%%%%%%%%%%%%%%%%%%%%%%%%%%%%%%%%%%%%%%%%%%%%%%%%%%%%%%%%%%%%%%%%%%%%%%%%%%%%%%%%%%%%%%%

Let us introduce the potential $v_i$ by $s_i=v_{i+h}-v_{i-n}$ so that  $v_{i}^{\prime}=\xi_i$. Therefore  the first relation in (\ref{pair}) becomes
\begin{equation}
\left(v_{i+h}-v_{i-n}\right)v_i^{\prime}=\left(v_{i+h+n}-v_i\right)v_{i+h}^{\prime}.
\label{111}
\end{equation}
It is worth remarking that the latter  is equivalent to the differential-difference equation
\begin{equation}
\prod_{j=1}^{h}v_{i+j-1}^{\prime}\cdot\prod_{j=1}^n\left(v_{i+j+h-1}-v_{i+j-n-1}\right)=1.
\label{3}
\end{equation}
In the case $h=1$, this class of equations were reported in \cite{Svinin2}.

Thus, we have in hand two classes of nonlinear differential-difference equations, namely, (\ref{u}) and (\ref{3}) which gives the compatibility of the linear equations (\ref{linear1}). These equations involve a pair of positive integers $(n, h)$, but one can see that to separate really different equations, one must suppose that $n$ and $h$ are co-prime positive integers. 

%%%%%%%%%%%%%%%%%%%%%%%%%%%%%%%%%%%%%%%%%%%%%%%%%%%%%%%%%%%%%%%%%%%%%%%%%%%%%%%%%%%%%%%%%%%%%%%%%%%%%%%%%%%%%%%%%%%%%%%%%%%%%%%%%%%%%%%%%%%%%%%%%%%%%%%%%%%%%%%%%%%%%%%%%%%%%%%%%%%%%%%%%%%%%%%%%%
\subsection{The third class of differential-difference equations}
%%%%%%%%%%%%%%%%%%%%%%%%%%%%%%%%%%%%%%%%%%%%%%%%%%%%%%%%%%%%%%%%%%%%%%%%%%%%%%%%%%%%%%%%%%%%%%%%%%%%%%%%%%%%%%%%%%%%%%%%%%%%%%%%%%%%%%%%%%%%%%%%%%%%%%%%%%%%%%%%%%%%%%%%%%%%%%%%%%%%%%%%%%%%%%%%%%

Let
\begin{equation}
T_{i+n}\equiv\prod_{j=1}^{n+h}u_{i+j-1}.
\label{ri2}
\end{equation}
By virtue of (\ref{xiu})
\begin{eqnarray}
T_{i+n}&=&\frac{\xi_{i}}{s_{i+n}}=\frac{\xi_{i+h}}{s_{i}} \label{ri} \\
&=&\frac{v_{i}^{\prime}}{s_{i+n}}=\frac{v_{i+h}^{\prime}}{s_{i}}. \label{ri1} 
\end{eqnarray}
In turn, by virtue of (\ref{ri1})  and equation (\ref{3})  
\[
\prod_{j=1}^{h}T_{i+j-1}=\prod_{j=1}^{n+h}\frac{1}{s_{i+j-n-1}}
\]
and
\begin{eqnarray}
\left(\prod_{j=1}^{h}T_{i+j-1}\right)^{\prime}&=&-\prod_{j=1}^{n+h}\frac{1}{s_{i+j-n-1}}\cdot\sum_{j=1}^{n+h}\frac{s_{i+j-n-1}^{\prime}}{s_{i+j-n-1}} \nonumber \\
&=&\prod_{j=1}^{h}T_{i+j-1}\left(\sum_{j=1}^{n+h}\frac{v_{i+j-2n-1}^{\prime}}{s_{i+j-n-1}} -\sum_{j=1}^{n+h}\frac{v_{i+j+h-n-1}^{\prime}}{s_{i+j-n-1}}\right). \nonumber
\end{eqnarray}
Using (\ref{ri1}) we get
\begin{eqnarray}
\left(\prod_{j=1}^{h}T_{i+j-1}\right)^{\prime}&=&\prod_{j=1}^{h}T_{i+j-1}\left(\sum_{j=1}^{n+h}T_{i+j-n-1}-\sum_{j=1}^{n+h}T_{i+j-1} \right) \nonumber \\
&=&\prod_{j=1}^{h}T_{i+j-1}\left(\sum_{j=1}^nT_{i+j-n-1}-\sum_{j=1}^nT_{i+j+h-1} \right). \label{INB}
\end{eqnarray}
One can see that (\ref{ri2}) relates two equations (\ref{INB}) and (\ref{u}).

%%%%%%%%%%%%%%%%%%%%%%%%%%%%%%%%%%%%%%%%%%%%%%%%%%%%%%%%%%%%%%%%%%%%%%%%%%%%%%%%%%%%%%%%%%%%%%%%%%%%%%%%%%%%%%%%%%%%%%%%%%%%%%%%%%%%%%%%%%%%%%%%%%%%%%%%%%%%%%%%%%%%%%%%%%%%%%%%%%%%%%%%%%%%%%%%%%%
\subsection{Linear equations on new wave function}
%%%%%%%%%%%%%%%%%%%%%%%%%%%%%%%%%%%%%%%%%%%%%%%%%%%%%%%%%%%%%%%%%%%%%%%%%%%%%%%%%%%%%%%%%%%%%%%%%%%%%%%%%%%%%%%%%%%%%%%%%%%%%%%%%%%%%%%%%%%%%%%%%%%%%%%%%%%%%%%%%%%%%%%%%%%%%%%%%%%%%%%%%%%%%%%%%%%

Let us introduce $\gamma_i$ such that 
\begin{equation}
\frac{\gamma_{i+1}}{\gamma_i}=\frac{1}{u_{i}}.
\label{quot}
\end{equation}
Clearly, that, by virtue of (\ref{xiu}) and (\ref{ri2}),
\begin{equation}
\frac{\gamma_{i+n}}{\gamma_i}=\frac{1}{\xi_i},\;\;
\frac{\gamma_{i+h}}{\gamma_{i}}=s_{i},\;\;
\frac{\gamma_{i+n+h}}{\gamma_{i}}=T_{i+n}. \label{sin}
\end{equation}
Let  $\phi_i\equiv \gamma_i\psi_i$. Linear equations  (\ref{linear1}) in terms of new wave function $\psi=\psi_i$ become
\begin{equation}
z\psi_{i+n}+T_{i+n}\psi_i=z\psi_{i+n+h},\;\;\;
\psi_i^{\prime}=z\psi_{i+n}-\frac{\gamma_i^{\prime}}{\gamma_i}\psi_i.
\label{linear2}
\end{equation}
With (\ref{sin}) and the second equation in (\ref{pair}), we have the following:
\[
\left(\frac{\gamma_{i+h}}{\gamma_i}\right)^{\prime}=\frac{\gamma_{i+h}}{\gamma_i}\left(\frac{\gamma_{i+h}^{\prime}}{\gamma_{i+h}
}-\frac{\gamma_i^{\prime}}{\gamma_i}\right)=s_{i}^{\prime}=\xi_{i+h}-\xi_{i-n}
\]
and then taking into account (\ref{ri}) we get
\begin{eqnarray}
\frac{\gamma_{i+h}^{\prime}}{\gamma_{i+h}}-\frac{\gamma_i^{\prime}}{\gamma_i}&=&\frac{\xi_{i+h}-\xi_{i-n}}{s_{i}}\nonumber \\
&=&\frac{\xi_{i+h}}{s_{i}}-\frac{\xi_{i+h-n}}{s_{i-n}} \nonumber \\
&=&T_{i+n}-T_i. \nonumber
\end{eqnarray}
We can resolve the latter as
\begin{equation}
\sum_{j=1}^{h}\frac{\gamma_{i+j-1}^{\prime}}{\gamma_{i+j-1}}=\sum_{j=1}^nT_{i+j-1}.
\label{4}
\end{equation}

%%%%%%%%%%%%%%%%%%%%%%%%%%%%%%%%%%%%%%%%%%%%%%%%%%%%%%%%%%%%%%%%%%%%%%%%%%%%%%%%%%%%%%%%%%%%%%%%%%%%%%%%%%%%%%%%%%%%%%%%%%%%%%%%%%%%%%%%%%%%%%%%%%%%%%%%%%%%%%%%%%%%%%%%%%%%%%%%%%%%%%%%%%%%%%%%%%
\section{Darboux transformation}
%%%%%%%%%%%%%%%%%%%%%%%%%%%%%%%%%%%%%%%%%%%%%%%%%%%%%%%%%%%%%%%%%%%%%%%%%%%%%%%%%%%%%%%%%%%%%%%%%%%%%%%%%%%%%%%%%%%%%%%%%%%%%%%%%%%%%%%%%%%%%%%%%%%%%%%%%%%%%%%%%%%%%%%%%%%%%%%%%%%%%%%%%%%%%%%%%%

\label{sec:3}

%%%%%%%%%%%%%%%%%%%%%%%%%%%%%%%%%%%%%%%%%%%%%%%%%%%%%%%%%%%%%%%%%%%%%%%%%%%%%%%%%%%%%%%%%%%%%%%%%%%%%%%%%%%%%%%%%%%%%%%%%%%%%%%%%%%%%%%%%%%%%%%%%%%%%%%%%%%%%%%%%%%%%%%%%%%%%%%%%%%%%%%%%%%%%%%%%%
\subsection{Quadratic discrete equation}
%%%%%%%%%%%%%%%%%%%%%%%%%%%%%%%%%%%%%%%%%%%%%%%%%%%%%%%%%%%%%%%%%%%%%%%%%%%%%%%%%%%%%%%%%%%%%%%%%%%%%%%%%%%%%%%%%%%%%%%%%%%%%%%%%%%%%%%%%%%%%%%%%%%%%%%%%%%%%%%%%%%%%%%%%%%%%%%%%%%%%%%%%%%%%%%%%%

Let us discuss Darboux transformation for linear equations (\ref{linear1}). We consider linear transformation in the form
\begin{equation}
\bar{\phi}_i=\phi_{i+h}+g_i\phi_i
\label{10}
\end{equation}
with some coefficient $g_i$ to be defined by condition that (\ref{10}) should be Darboux transformation for (\ref{linear1}). Consider the transformation of the first equation in (\ref{linear1}). We have
\begin{eqnarray}
z\bar{s}_i\left(\phi_{i+n+h}+g_{i+n}\phi_{i+n}\right)+\phi_{i+h}+g_i\phi_i&=&z\left(\phi_{i+h-n}+g_{i+n+h}\phi_{i+n+h}\right)\nonumber\\
                                                                          &=&zs_{i+h}\phi_{i+n+h}+\phi_{i+h}+zg_{i+n+h}\phi_{i+n+h}\nonumber
\end{eqnarray}
and therefore
\[
z\left(\bar{s}_i-s_{i+h}-g_{i+n+h}\right)\phi_{i+n+h}+g_i\phi_i+z\bar{s}_ig_{i+n}\phi_{i+n}=0.
\]
Requiring that (\ref{10}) to be Darboux transformation gives the relations
\[
g_{i+n+h}-g_i=\bar{s}_i-s_{i+h}=\bar{v}_{i+n+h}-\bar{v}_i+v_{i+h}-v_{i+n+2h},\;\;\;
g_is_i=\bar{s}_ig_{i+n}
\]
the first of which is solved by $g_i=\bar{v}_i-v_{i+h}$ and therefore the second one is equivalent to the following discrete equation:
\begin{equation}
\left(\bar{v}_i-v_{i+h}\right)\left(v_{i+n+h}-v_i\right)=\left(\bar{v}_{i+n+h}-\bar{v}_i\right)\left(\bar{v}_{i+n}-v_{i+n+h}\right).
\label{2}
\end{equation}
Note that this equation in the special case $h=1$ has appeared in \cite{Svinin2}. One can check, that it can be also written as
\begin{equation}
\left(v_{i+n+h}-v_i\right)\left(\bar{v}_{i+n+h}-v_{i+h}\right)=\left(\bar{v}_{i+n+h}-\bar{v}_i\right)\left(\bar{v}_{i+n}-v_i\right)
\label{21}
\end{equation}
and in the form
\begin{equation}
\left(\bar{v}_{i+n}-v_i\right)\left(\bar{v}_i-v_{i+h}\right)=\left(\bar{v}_{i+n+h}-v_{i+h}\right)\left(\bar{v}_{i+n}-v_{i+n+h}\right).
\label{22}
\end{equation}
We observe that replacing $v_i\leftrightarrow \bar{v}_{i}$ and $(n, h)\leftrightarrow (h, n)$ in (\ref{2}), we obtain (\ref{21}). This means that two different pairs of parameters $(n, h)$ and $(h, n)$correspond in fact to the same equation. Thus except the case $(n, h)=(1, 1)$, we can assume, without loss of generality, that $n$ and $h$ are co-prime positive integer numbers with condition $h<n$.  

Making use of (\ref{22}), we observe that this quadratic equation has the following integral:
\begin{equation}
I_i=\prod_{j=1}^ng_{i+j-1}\cdot\prod_{j=1}^{h}h_{i+j-1},
\label{Ii}
\end{equation}
where $h_i\equiv g_{i+n}+s_i=\bar{v}_{i+n}-v_i$. We can also present $I_i$  in the form
\[
I_i=\prod_{j=1}^{n+h}\left(\bar{v}_{i+j-1}-v_{i+j+h-1}\right).
\]
The latter needs some explanation. This formula involve $\bar{v}_{i+\alpha}$ with $\alpha\in\{0,\ldots, n+h-1\}$, while $\alpha$ in $v_{i+\alpha}$ is calculated modulo $n+h$.

Remark that the equation $I_i=c$ with some constant $c$ in simplest case $(n, h)=(1, 1)$ is  the lpKdV equation (\ref{lpkdV}) \cite{Nijhoff1, Nijhoff2} also known as $H_1$ equation in the ABS classification of quad-equations satisfying some symmetry conditions and sharing the property of three-dimensional consistency \cite{Adler}. 

One can consider the relation 
\begin{equation}
\prod_{j=1}^{n+h}\left(\bar{v}_{i+j-1}-v_{i+j+h-1}\right)=c
\label{gen-lpkdV}
\end{equation}
with an arbitrary constant $c$ in a sense as a generalization of the lpKdV equation (\ref{lpkdV}). We have presented these equations in the particular case $h=1$ in \cite{Svinin2}.
Clearly, they are  higher order partial difference equations of the form (\ref{form}) with $p=n+h$. 

\subsection{Discrete zero-curvature representation for the equation (\ref{2})}

Let $\phi_{k,i}\equiv \phi_{i+k-1}$ for $k=1,\ldots, n+h$ and $\Phi_i\equiv (\phi_{1,i},\ldots, \phi_{n+h,i})^T$. We can rewrite (\ref{10}) in matrix form $\bar{\Phi}_i=V_i\Phi_i$ or more explicitly as
\[
\bar{\phi}_{1, i}=g_i\phi_{1,i}+\phi_{h+1, i},\ldots,
\bar{\phi}_{n, i}=g_{i+n-1}\phi_{n,i}+\phi_{n+h, i},
\]
\[
\bar{\phi}_{n+1, i}=h_i\phi_{n+1,i}+\frac{1}{z}\phi_{1,i},\ldots,
\bar{\phi}_{n+h, i}=h_{i+h-1}\phi_{n+h, i}+\frac{1}{z}\phi_{h, i}.
\]
Obviously, the second equation which complete discrete zero-curvature representation for (\ref{2}) being of the form $\Phi_{i+1}=U_i\Phi_i$  is explicitly given by the equations
\[
\phi_{k,i+1}=\phi_{k+1,i}\;\;\;\mbox{for}\;\;\;k=1,\ldots, n+h-1,\;\;
\phi_{n+h,i+1}=\frac{1}{z}\phi_{1,i}+s_{i+n}\phi_{n+1,i}.
\]
Thus the discrete zero-curvature representation for quadratic equation (\ref{2}) is given by matrix equation $V_{i+1}U_i=\bar{U}_iV_i$.

%%%%%%%%%%%%%%%%%%%%%%%%%%%%%%%%%%%%%%%%%%%%%%%%%%%%%%%%%%%%%%%%%%%%%%%%%%%%%%%%%%%%%%%%%%%%%%%%%%%%%%%%%%%%%%%%%%%%%%%%%%%%%%%%%%%%%%%%%%%%%%%%%%%%%%%%%%%%%%%%%%%%%%%%%%%%%%%%%%%%%%%%%%%%
\subsection{An example. Zero-curvature representation for lpKdV}
%%%%%%%%%%%%%%%%%%%%%%%%%%%%%%%%%%%%%%%%%%%%%%%%%%%%%%%%%%%%%%%%%%%%%%%%%%%%%%%%%%%%%%%%%%%%%%%%%%%%%%%%%%%%%%%%%%%%%%%%%%%%%%%%%%%%%%%%%%%%%%%%%%%%%%%%%%%%%%%%%%%%%%%%%%%%%%%%%%%%%%%%%%%%

Consider the simplest case $(n, h)=(1, 1)$ for which we have the following pair of auxiliary linear equations:
\[
\left(
\begin{array}{c}
\phi_{1,i+1} \\
\phi_{2,i+1}
\end{array}
\right)=
\left(
\begin{array}{cc}
0   &\;\; 1 \\
\displaystyle
\frac{1}{z} &\;\; v_{i+2}-v_i
\end{array}
\right)
\left(
\begin{array}{c}
\phi_{1,i} \\
\phi_{2,i}
\end{array}
\right)
\]
and
\[
\left(
\begin{array}{c}
\bar{\phi}_{1,i} \\
\bar{\phi}_{2,i}
\end{array}
\right)=
\left(
\begin{array}{cc}
\bar{v}_i-v_{i+1}   & 1 \\
\displaystyle
\frac{1}{z} & \bar{v}_{i+1}-v_i
\end{array}
\right)
\left(
\begin{array}{c}
\phi_{1,i} \\
\phi_{2,i}
\end{array}
\right).
\]
Let $\varphi_{1,i}=\phi_{1,i}$ and $\varphi_{2,i}=\left(v_{i+1}-v_i\right)\phi_{1,i}-\phi_{2,i}$. In terms of these new wave functions we have
\begin{equation}
\left(
\begin{array}{c}
\varphi_{1,i+1} \\
\varphi_{2,i+1}
\end{array}
\right)=
\left(
\begin{array}{cc}
v_{i+1}-v_i   &\;\;\; -1 \\
\displaystyle
-(v_{i+1}-v_i)^2-\frac{1}{z} &\;\;\; v_{i+1}-v_i
\end{array}
\right)
\left(
\begin{array}{c}
\varphi_{1,i} \\
\varphi_{2,i}
\end{array}
\right)
\label{zcr}
\end{equation}
and
\[
%\fl
\left(
\begin{array}{c}
\bar{\varphi}_{1,i} \\
\bar{\varphi}_{2,i}
\end{array}
\right)=
\left(
\begin{array}{cc}
\bar{v}_i-v_i   &\;\;\; -1 \\
\displaystyle
\left(\bar{v}_i-v_i\right)\left(\bar{v}_{i+1}-\bar{v}_i\right)+\left(v_i-\bar{v}_{i+1}\right)\left(v_{i+1}-v_i\right)-\frac{1}{z} &\;\;\; \bar{v}_i-v_i
\end{array}
\right)
\left(
\begin{array}{c}
\varphi_{1,i} \\
\varphi_{2,i}
\end{array}
\right).
\]
If we make use of (\ref{lpkdV}) we obtain
\[
\left(
\begin{array}{c}
\bar{\varphi}_{1,i} \\
\bar{\varphi}_{2,i}
\end{array}
\right)=
\left(
\begin{array}{cc}
\bar{v}_i-v_i   &\;\;\; -1 \\
\displaystyle
-(\bar{v}_i-v_i)^2+c-\frac{1}{z} &\;\;\; \bar{v}_i-v_i
\end{array}
\right)
\left(
\begin{array}{c}
\varphi_{1,i} \\
\varphi_{2,i}
\end{array}
\right).
\]
One can see that the latter is more like (\ref{zcr}). Let $c=\alpha-\beta$ and $1/z=\alpha-\lambda$, where $\lambda$ is a new ``spectral'' parameter. Then, as a result we obtain well-known symmetric zero-curvature representation defined by a pair of equations
\[
\left(
\begin{array}{c}
\varphi_{1,i+1} \\
\varphi_{2,i+1}
\end{array}
\right)=
\left(
\begin{array}{cc}
v_{i+1}-v_i   &\;\;\; -1 \\
\displaystyle
-(v_{i+1}-v_i)^2+\lambda-\alpha &\;\;\; v_{i+1}-v_i
\end{array}
\right)
\left(
\begin{array}{c}
\varphi_{1,i} \\
\varphi_{2,i}
\end{array}
\right)
\]
and
\[
\left(
\begin{array}{c}
\bar{\varphi}_{1,i} \\
\bar{\varphi}_{2,i}
\end{array}
\right)=
\left(
\begin{array}{cc}
\bar{v}_i-v_i   &\;\;\; -1 \\
\displaystyle
-(\bar{v}_i-v_i)^2+\lambda-\beta &\;\;\; \bar{v}_i-v_i
\end{array}
\right)
\left(
\begin{array}{c}
\varphi_{1,i} \\
\varphi_{2,i}
\end{array}
\right).
\]

%%%%%%%%%%%%%%%%%%%%%%%%%%%%%%%%%%%%%%%%%%%%%%%%%%%%%%%%%%%%%%%%%%%%%%%%%%%%%%%%%%%%%%%%%%%%%%%%%%%%%%%%%%%%%%%%%%%%%%%%%%%%%%%%%%%%%%%%%%%%%%%%%%%%%%%%%%%%%%%%%%%%%%%%%%%%%%%%%%%%%%%%%%%%
\section{Generalized symmetries}
%%%%%%%%%%%%%%%%%%%%%%%%%%%%%%%%%%%%%%%%%%%%%%%%%%%%%%%%%%%%%%%%%%%%%%%%%%%%%%%%%%%%%%%%%%%%%%%%%%%%%%%%%%%%%%%%%%%%%%%%%%%%%%%%%%%%%%%%%%%%%%%%%%%%%%%%%%%%%%%%%%%%%%%%%%%%%%%%%%%%%%%%%%%%%

\label{sec:4}

One knows that the quad-equations which are consistent around a the cube possess generalized symmetries \cite{Levi1, Levi2, Levi3}, \cite{Rasin}, \cite{Tongas}. In this section we derive the integrable hierarchies of evolution equations on the field $v_i$ which apparently are symmetries for discrete equations (\ref{2}). 

It was shown in \cite{Svinin1}, that linear equation (\ref{linear2}) together with 
\[
\partial_s\psi_i=z^{sh}\psi_{i+shn}+\sum_{j=1}^{sh}(-1)^jz^{sh-j}S_{sn}^{j}(i-(j-1)n)\psi_{i+(sh-j)n}
\]
constitute Lax pair for hierarchy of evolution equations\footnote{In the following we get rid of the factor $(-1)^{sh}$ by replacing $t_s\mapsto (-1)^{sh}t_s$.}
\begin{equation}
\partial_sT_i=(-1)^{sh}T_i\left\{S_{sn}^{sh}(i-shn+n+h)-S_{sn}^{sh}(i-shn)\right\},
\label{T}
\end{equation}
where discrete polynomial $S^k_r$ attached to some pair $(n, h)$ is defined by
\[
S^k_r(i)\equiv\sum_{\{\lambda_j\}\in B_{k, r}} T_{i+\lambda_1h+(k-1)n}\cdots T_{i+\lambda_{k-1}h+n}T_{i+\lambda_kh},
\]
with $B_{k, r}\equiv \{\lambda_j : 0\leq \lambda_1\leq\cdots\leq\lambda_k\leq r-1\}$. Suppose that 
\begin{equation}
\partial_s u_i=u_i\left\{\omega(i)-\omega(i+1)\right\},
\label{U}
\end{equation}
with some  discrete polynomial $\omega$ in $u$, such that (\ref{ri2}) should give Miura-type transformation relating (\ref{U}) to (\ref{T}). Direct inspection gives $\omega(i)=S_{sn}^{sh}(i-shn+n)$ 
where $T_i$ is replaced by $u_i$ by virtue of (\ref{ri2}). 

By virtue of (\ref{quot}) we get $\partial_s\gamma_i/\gamma_i=\omega_i$ and
\[
\partial_s\phi_i=z^{sh}v^{[sh]}_i\phi_{i+shn}+\sum_{j=1}^{sh-1}(-1)^jz^{sh-j}v^{[sh-j]}_iS_{sn}^{j}(i-(j-1)n)\phi_{i+(sh-j)n}
\]
where $v^{[r]}_i\equiv v_i^{\prime}v_{i+n}^{\prime}\cdots v_{i+(r-1)n}^{\prime}$ for any positive integer $r$. We get as a result compatibility condition for pair of this evolution linear equation with the first one in (\ref{linear1}) in the form
\begin{equation}
\partial_sv_i=v^{\prime}_iS_{sn}^{sh-1}(i-shn+2n).
\label{ierarh0}
\end{equation}
Let us discuss the latter. 
Our assumption is that in general case the right-hand side of (\ref{ierarh0}), by virtue of (\ref{3}) and (\ref{ri1}),  is some discrete polynomial in $c_i\equiv s_i^{-1}$, which in the following we denote as $P^{(n, h)}_s=P^{(n, h)}_s[c]$.
Moreover, actual calculations show that two pairs $(n, h)$ and $(h, n)$ correspond in fact to the same hierarchy, or more exactly that $P^{(n, h)}_s=P^{(h, n)}_s$, and we accept this as the second hypothesis. It is the more natural if we assume that (\ref{ierarh0}) yields generalized symmetries for equation of the form (\ref{gen-lpkdV}) and this is our the third hypothesis. 

We know that in the particular case $h=1$,  $v^{\prime}_i$ is expressed via $v_i$ as $v^{\prime}_i=\prod_{j=1}^nc_{i+j-1}$  and in this case
\begin{eqnarray}
\partial_sv_i&=&P^{(n, 1)}_s\nonumber\\
             &=&S_{sn}^{s-1}(i-(s-2)n)\prod_{j=1}^nc_{i+j-1}.\label{Hierar}
\end{eqnarray}
This hierarchy was presented in \cite{Svinin2}. In the more particular case $n=1$, (\ref{Hierar}) gives generalized symmetries for the lpKdV equation (\ref{lpkdV}) in explicit form. 

Finally, let us write down discrete polynomials in $c_i$ defining the first flows in (\ref{ierarh0}) and corresponding to some pairs of co-prime numbers $(n, h)$ such that $h<n$, for $h=1, 2, 3$. We get the following.
\[
P^{(n, 1)}_1(i)=\prod_{j=1}^nc_{i+j-1},
\]
\[
P^{(3, 2)}_1(i)=c_{i-1}c_{i}c_{i+1}c_{i+2}+c_{i}c_{i+1}\left\{c_{i-3}c_{i-1}+c_{i+2}c_{i+4}\right\},
\]
\begin{eqnarray}
P^{(5, 2)}_1(i)&=&c_{i-1}c_{i}c_{i+1}c_{i+2}c_{i+3}c_{i+4}+c_{i}c_{i+1}c_{i+2}c_{i+3}\left\{c_{i-3}c_{i-1}+c_{i+4}c_{i+6}\right\}\nonumber\\
&&+c_{i}c_{i+3}\left\{c_{i-5}c_{i-3}c_{i-1}c_{i+1}+c_{i+2}c_{i+4}c_{i+6}c_{i+8}\right\},\nonumber
\end{eqnarray}
\begin{eqnarray}
P^{(7, 2)}_1(i)&=&c_{i-1}c_{i}c_{i+1}c_{i+2}c_{i+3}c_{i+4}c_{i+5}c_{i+6}+c_{i}c_{i+1}c_{i+2}c_{i+3}c_{i+4}c_{i+5}\left\{c_{i-3}c_{i-1}+c_{i+6}c_{i+8}\right\}\nonumber\\
&&+c_{i}c_{i+2}c_{i+3}c_{i+5}\left\{c_{i-5}c_{i-3}c_{i-1}c_{i+1}+c_{i+4}c_{i+6}c_{i+8}c_{i+10}\right\}\nonumber\\
&&+c_{i}c_{i+5}\left\{c_{i-7}c_{i-5}c_{i-3}c_{i-1}c_{i+1}c_{i+3}+c_{i+2}c_{i+4}c_{i+6}c_{i+8}c_{i+10}c_{i+12}\right\},\nonumber
\end{eqnarray}
\begin{eqnarray}
P^{(4, 3)}_1(i)&=&c_{i-2}c_{i-1}c_{i}c_{i+1}c_{i+2}c_{i+3}+c_{i-4}c_{i-1}c_{i}c_{i+1}c_{i+2}c_{i+5}\nonumber\\
&&+c_{i-1}c_{i}c_{i+1}c_{i+2}\left\{c_{i-4}c_{i-2}+c_{i+3}c_{i+5}\right\}\nonumber\\
&&+c_{i-2}c_{i}c_{i+1}c_{i+3}\left\{c_{i-5}c_{i-1}+c_{i+2}c_{i+6}\right\}\nonumber\\
&&+c_{i}c_{i+1}\left\{c_{i-8}c_{i-5}c_{i-4}c_{i-2}+c_{i-5}c_{i-4}c_{i-2}c_{i-1}+c_{i+2}c_{i+3}c_{i+5}c_{i+6}+c_{i+3}c_{i+5}c_{i+6}c_{i+9}\right\}.\nonumber
\end{eqnarray}

%%%%%%%%%%%%%%%%%%%%%%%%%%%%%%%%%%%%%%%%%%%%%%%%%%%%%%%%%%%%%%%%%
\section{Conclusion}
%%%%%%%%%%%%%%%%%%%%%%%%%%%%%%%%%%%%%%%%%%%%%%%%%%%%%%%%%%%%%%%%%

In the paper we have presented a class of partial difference equations parametrized by pairs of co-prime numbers $(n, h)$. They are given, on the one hand by quadratic equation (\ref{2}) and from the other hand by lpKdV-type equations (\ref{gen-lpkdV}). 

There remain some problems. The first one: are these systems satisfy the condition of three-dimensional consistency? It is well-known that it is valid in the case $(n, h)=(1, 1)$, that is, for the lpKdV equation (\ref{lpkdV}) \cite{Adler}. Using of the test proposed in \cite{Adler1} for other cases  of (\ref{gen-lpkdV}) runs into insurmountable technical difficulties especially that one have to prove this property for an infinite number of equations. 

In addition, we did not strictly prove that (\ref{ierarh0}) yields generalized symmetries for (\ref{gen-lpkdV}), but we hope to fill these gaps in the future.


\begin{thebibliography}{00}

\bibitem{Adler}
V.E. Adler, A.I. Bobenko and Yu. B. Suris,
Classification of integrable equations on quad-graphs. The consistency approach,
\textit{Commun. Math. Phys.} \textbf{233} (2003) 513--543.


\bibitem{Adler1}
V.E. Adler and V.V. Postnikov, 
On discrete 2D integrable equations of higher order,
\textit{J. Phys. A.: Math. Theor.} \textbf{47} (2014) Art. No. 045206.


\bibitem{Bogoyavlenskii}
O.I. Bogoyavlenskii,
Some constructions of integrable dynamical systems,
\textit{Math. USSR-Izv.} \textbf{31} (1988) 47--76.

\bibitem{Hirota}
R. Hirota and J. Satsuma,
Nonlinear partial difference equations. I. A difference analogue of the Korteweg-de Vries equation,
\textit{J.  Phys. Soc.  Japan}  \textbf{43} 1424--1433. 

\bibitem{Itoh}
Y. Itoh,
An $H$-theorem for a system of competing species, 
\textit{Proc. Japan Acad.} \textbf{51} (1975) 374--379.

\bibitem{Levi1}
D. Levi, M. Petrera and C. Scimiterna,
The lattice Schwarzian KdV equation and its symmetries ,
\textit{J. Phys. A: Math. Theor.} \textbf{40} (2007) 12753--12761.

\bibitem{Levi2}
D. Levi, M. Petrera, 
Continuous symmetries of the lattice potential KdV equation,
\textit{J. Phys. A: Math. Theor.} \textbf{40} (2007) 4141--4159.

\bibitem{Levi3}
D. Levi, M. Petrera, C. Scimiterna and R. Yamilov,
On Miura Transformations and Volterra-Type Equations Associated with the Adler-Bobenko-Suris Equations,
\textit{SIGMA} \textbf{4} (2008) Paper 077.  



\bibitem{Manakov}
S.V. Manakov,
Complete integrability and stochastization of discrete dynamical systems,
\textit{Sov. Phys.-JETP} \textbf{40} (1975) 269--274.


\bibitem{Narita}
K. Narita, 
Soliton solutions to extended Volterra equation,
 \textit{J. Phys. Soc. Japan} \textbf{51} (1982) 1682--1685.

\bibitem{Nijhoff1}
F.W. Nijhoff, G.R.W. Quispel and H.W. Capel, 
Direct linearization of nonlinear difference-difference equations
\textit{Phys. Lett. A} \textbf{97} (1983) 125--128.

\bibitem{Nijhoff2}
F.W. Nijhoff and H.W. Capel, 
The discrete Korteweg-de Vries equation, 
\textit{Acta Appl. Math.} \textbf{39} (1995) 133--158.

\bibitem{Rasin}
O.G. Rasin and P.E. Hydon,
Symmetries of integrable difference equations on the quad-graph,
\textit{Stud. Appl. Math.} \textbf{119} (2007) 253--269.


\bibitem{Svinin1}
A.K. Svinin,
On some class of homogeneous polynomials and explicit form of integrable hierarchies of differential-difference equations, 
\textit{J. Phys. A: Math. Theor.} \textbf{44} (2011) Art. No. 165206.

\bibitem{Svinin2}
A.K. Svinin,
On some integrable lattice related by the Miura-type transformation to the Itoh-Narita-Bogoyavlenskii lattice,
\textit{J. Phys. A: Math. Theor.} \textbf{44} (2011) Art. No. 465210.

\bibitem{Svinin3}
A.K. Svinin, 
On some classes of discrete polynomials and ordinary difference equations,
\textit{J. Phys. A: Math. Theor.} \textbf{47} (2014) Art. No. 155201.

\bibitem{Tongas}
A. Tongas, D. Tsoubelis and P. Xenitidis,
Affine linear and $D_4$ symmetric lattice equations: symmetry analysis and reductions,
\textit{J. Phys. A: Math. Theor.} \textbf{40} (2007) 13353--13384.

\bibitem{Volterra}
V. Volterra
\textit{Le\c{c}on sur la th\'eorie math\'ematique de la lutte pour la vie} 
Gauthier-Villars, Paris, 1931. 

\bibitem{Yamilov}
R.I. Yamilov, 
Classification of discrete evolution equations, 
\textit{Usp. Mat. Nauk} \textbf{38} (1983) 155--156 (in Russian).

\end{thebibliography}
\end{document}